\documentclass[Times,4pt,aps,pra,twocolumn,showpacs,amsmath,amssymb,floatfix,footinbib,superscriptaddress]{revtex4-2}
\usepackage{amsmath,natbib,graphicx,subfigure,amssymb,graphics,amsmath,mathrsfs,CJK,color,comment}
\usepackage{multirow,fancyhdr,color,bm,tabularx,psfrag,geometry,dcolumn,datetime,physics,booktabs}
\usepackage[colorlinks=true,linkcolor=blue,urlcolor=blue,citecolor=blue]{hyperref}
\usepackage[mathlines]{lineno}
\def\footnoterule{\kern -1mm \hrule width 5.8cm \kern 2.2mm}
\usepackage{tikz,xcolor,hyperref}
\definecolor{lime}{HTML}{A6CE39}
\DeclareRobustCommand{\orcidicon}{%
    \begin{tikzpicture}
    \draw[lime, fill=lime] (0,0)
    circle [radius=0.16]
    node[white] {{\fontfamily{qag}\selectfont \tiny ID}};\draw[white, fill=white] (-0.0625,0.095)
    circle [radius=0.007];
    \end{tikzpicture}
    \hspace{-2mm}}
\foreach \x in {A, ..., Z}
{\expandafter\xdef\csname orcid\x\endcsname{\noexpand\href{https://orcid.org/\csname orcidauthor\x\endcsname}{\noexpand\orcidicon}}}

\begin{document}
\title{Negative refraction with low absorption using EIT in a four-level left-handed atomic system}
\author{Shun-Cai Zhao\orcidA{}}
\email[Corresponding author: ]{zhaosc@kust.edu.cn.}
\affiliation{Physics department, Kunming University of Science and Technology, Kunming, 650093,PR China} 
\author{Zheng-Dong Liu}
\affiliation{Engineering Research Center for Nanotechnology,Nanchang University,Nanchang 330047,PR China}
\affiliation{Institute of Modern Physics,Nanchang University,Nanchang 330031,PR China}
 \author{ Gen Li}
\affiliation{Engineering Research Center for Nanotechnology,Nanchang University,Nanchang 330047,PR China}
\affiliation{Institute of Modern Physics,Nanchang University,Nanchang 330031,PR China}
\author{ Nian Liu}
\affiliation{Engineering Research Center for Nanotechnology,Nanchang University,Nanchang 330047,PR China}
\affiliation{Institute of Modern Physics,Nanchang University,Nanchang 330031,PR China}
\begin{abstract}
We suggest a scheme for obtaining negative refraction with low
absorption in a left-handed atomic system.Under the the appropriate
conditions,the atomic system displays negative refraction with
negative permittivity and permeability(Left-handedness)in a common
frequency range,simultaneously.And the imaginary parts of
permittivity and permeability show transparently propagate in the
same frequency range.Finally,the negative refraction show low
absorption due to the EIT effect,and the figure of merit
demonstrated this in this resonant atomic system.
\begin{description}
\item[PACs]{42.50.Gy}
\item[Keywords]{negative refraction; low absorption; electromagnetic induced transparency(EIT); Left-handedness}
\end{description}
\end{abstract}
\maketitle

\section{Introduction}
Negative refraction of light,first predicted to occur in materials
with simultaneous negative permittivity and permeability in 1968[1],
has attracted considerable attention in the last decade.These
materials with negative refraction index promise many surprising and
even counterintuitive electromagnetical and optical effects,such as
the reversals of both Doppler shift and Cherenkov effect,negative
refraction[1],amplification of evanescent waves and subwavelength
focusing [2-4], negative Goos-H$\ddot{a}$nchen shift[5]. One of the
key practical application for these materials was discovered in the
year 2000 when Pendry demonstrated that a slab with a negative index
of refraction can image objects with, in principle, unlimited
resolution [2].Since Pendry's suggestion, the interest in these
materials has been continuously growing and there have been a large
number of theoretical developments and experimental advances [6-15].
A key difficulty of these experiments, which is particularly
pronounced in the optical domain, is the large absorption that
accompanies negative refraction. The performance of the left-handed
materials is typically characterized by the figure of merit F =
-Re(n)/$|Im(n)|$[16-17]. For all recent experiments in the optical
region, the figure of merit is of order unity, F$\approx$1, which is
a key limitation for many potential applications. It therefore
remains a big challenge to obtain negative refraction with low
absorption in the optical region of the spectrum.

Thus,the realization of negative refraction material with low
absorption is of great significance.And some effort[17-20]has been
done to realize negative refraction with low absorption.Ref.[18]
realized negative refraction with strongly suppressed absorption by
using a chiral medium due to quantum interference effects similar to
electromagnetically induced transparency (EIT)[21-22].Negative
refraction with reduced absorption due to destructive quantum
interference in coherently driven atomic media was also discussed in
Ref.[18].The negative refraction without simultaneously requiring
both negative electric permittivity and magnetic permeability [1]
demonstrates deeply depressing absorption in Ref.[19]. Ref.[19] gets
this at the ideal situation that the two chirality coefficients have
the same amplitude but the opposite phase.A pair of electric-dipole
Raman transitions and utilizing magneto-electric cross coupling to
achieve a negative index of refraction with low absorption is
obtained in Ref.[20].

In this paper we propose a scheme to realized negative refractive
index with low absorption in the left-handed atomic system using by
the EIT effect.The atomic system shows negative refraction with
simultaneously negative permittivity and negative
permeability(Left-handednes),and the transparent propagation of the
probe field demonstrated by the imaginary parts of the permittivity
and permeability brings about the low absorption of the negative
refraction in such atomic system.

The paper is organized as follows: Section 2 establishes the model,
and the evolution equations of the atomic system assuming the dipole
approximation and the rotating wave approximation. Section 3 is
devoted to present the numerical results and to discuss the origin
of low absorption. Finally, Sec.4 summarizes the conclusions.

\section{Theoretical model}

Consider a four-level quasi-$\Lambda$ atomic ensemble interacting
with two coherent optical fields,i.e.the coupling beam (with
frequency $\omega_{c}$)and probe light field(with frequency
$\omega_{p}$). The atomic configuration is schematically shown in
figure 1. The two levels, $|2\rangle$ and $|4\rangle$ have the same
parity and
$\mu_{42}=\langle4|\vec{{\mu_{42}}}|2\rangle$$\neq0$,where
$\vec{{\mu_{42}}}$ is the magnetic dipole operator, and the levels
$|3\rangle$ and $|4\rangle$ have an odd parity with
$d_{43}=\langle3|\vec{d_{43}}|4\rangle$$\neq0$,where $\vec{d_{43}}$
is the electric dipole operator.The coupling field drivers the
transition from the upper level $|3\rangle$ to a pair of ground
levels $|1\rangle$(transition frequency $\omega_{31}$)with Rabi
frequency $\Omega_{1}$ ,and $|2\rangle$ (transition frequency
$\omega_{32}$)with Rabi frequency $\Omega_{2}$ .The probe field with
Rabi frequency $\Omega_{p}$ interacts with the transition
$|3\rangle$ to $|4\rangle$ (transition frequency $\omega_{34}$).And
its  electric and magnetic components couple the level pairs
$|3\rangle$-$|4\rangle$ and $|2\rangle$-$|4\rangle$, respectively.
Here we treat the Rabi frequency
$\Omega_{p}$($|\Omega_{p}|$$\ll$$|\Omega_{1,2}|$) as real parameter,
$\Omega_{1}$=$\Omega_{2}$ as complex parameters:$\Omega_{1,2}$=
$|\Omega_{1,2}|$ $e^{i\Phi}$,where $\Phi$ is the phase of the
coupling field. The frequency detunings of these optical fields are
$\Delta_{1}$,$\Delta_{2}$ and
$\Delta_{P}$,respectively.$\gamma_{i=1,2,3}$ are the population
decay rates from level $|3\rangle$ to $|1\rangle$,$|2\rangle$ and
$|4\rangle$,respectively.

\begin{figure}[h!]
  \centering
  \includegraphics[width=3.1in]{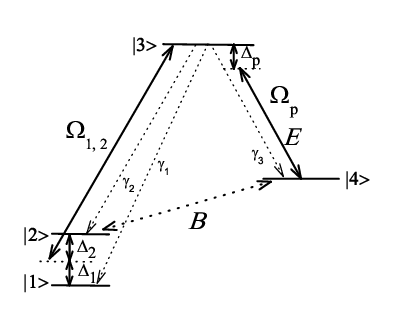}
  \hspace{0in}%
  \caption{Schematic diagram of four-level system
interacting with a coupling beam $\Omega_{1,2}$ and a probe field
$\Omega_{p}$.The level pairs $|3\rangle-|4\rangle$,
$|2\rangle-|4\rangle$ are coupled to the electric and magnetic
components of the probe light,respectively.}
\hspace{0in}%
\end{figure}\label{Fig.1}

Using the density-matrix approach, the time-evolution of the system
is described as
\begin{equation}
\frac{d\rho}{dt}=-\frac{i}{\hbar}[H,\rho]+\Lambda\rho ,
\end{equation}
Where $\Lambda\rho$ represents the irreversible decay part in the
system.Under the dipole approximation and the rotating wave
approximation the density matrix equations described the system are
written as follows:
\begin{equation}
\dot{\rho_{11}}=\gamma_{1}\rho_{33}+(-i\Omega^{\star}_{1}\rho_{31}+H.c.),
\end{equation}
\begin{equation}
\dot{\rho_{12}}=i[(\Delta_{1}-\Delta_{2})\rho_{12}+\Omega_{2}\rho_{13}-\Omega^{\star}_{1}\rho_{32}],
\end{equation}
\begin{equation}
\dot{\rho_{13}}=-\frac{1}{2}(\gamma_{1}+\gamma_{2}+\gamma_{3})\rho_{13}+i\Omega^{\star}_{1}(\rho_{11}-\rho_{33})+i(\Omega^{\star}_{2}\rho_{12}+\Delta_{1}\rho_{13}+\Omega^{\star}_{p}\rho_{14}),
\end{equation}
\begin{equation}
\dot{\rho_{14}}=i[(\Delta_{1}-\Delta_{p})\rho_{14}+\Omega_{p}\rho_{13}-\Omega^{\star}_{1}\rho_{34}],
\end{equation}
\begin{equation}
\dot{\rho_{22}}=\gamma_{2}\rho_{33}+(-i\Omega^{\star}_{2}\rho_{32}+H.c.),
\end{equation}
\begin{equation}
\dot{\rho_{23}}=-\frac{1}{2}(\gamma_{1}+\gamma_{2}+\gamma_{3})\rho_{23}+i\Omega_{1}(\rho_{22}-\rho_{33})+i(\Omega^{\star}_{1}\rho_{21}+\Delta_{2}\rho_{23}+\Omega^{\star}_{p}\rho_{24}),
\end{equation}
\begin{equation}
\dot{\rho_{24}}=i[(\Delta_{2}-\Delta_{p})\rho_{24}+\Omega_{p}\rho_{23}-\Omega^{\star}_{2}\rho_{24}],
\end{equation}
\begin{equation}
\dot{\rho_{33}}=-(\gamma_{1}+\gamma_{2}+\gamma_{3})\rho_{33}+(i\Omega^{\star}_{1}\rho_{31}+i\Omega^{\star}_{2}\rho_{32}+i\Omega^{\star}_{p}\rho_{34}+H.c.),
\end{equation}
\begin{equation}
\dot{\rho_{34}}=-\frac{1}{2}(\gamma_{1}+\gamma_{2}+\gamma_{3})\rho_{34}-i\Omega_{p}(\rho_{44}-\rho_{33})-i(\Omega_{1}\rho_{14}+\Omega_{2}\rho_{24})
-i\Delta_{p}\rho_{34}
\end{equation}
where$\Delta_{1}$=$\omega_{c}$-$\omega_{31}$,$\Delta_{2}$=$\omega_{c}$-$\omega_{32}$
and $\Delta_{P}$=$\omega_{p}$-$\omega_{34}$. And the above density
matrix elements obey the
conditions:$\rho_{11}+\rho_{22}+\rho_{33}+\rho_{44}$=1 and
$\rho_{ij}$=$\rho_{ji}^{\ast}$.In the following,we will discuss the
electric and magnetic responses of the medium to the probe
field.When discussing how the detailed properties of the atomic
transitions between the levels are related to the electric and
magnetic susceptibilities, one must make a distinction between
macroscopic fields and the microscopic local fields acting upon the
atoms in the vapor.In a dilute vapor,there is little difference
between the macroscopic fields and the local fields that act on any
atoms(molecules or group of molecules)[23].But in dense media with
closely packed atoms(molecules),the polarization of neighboring
atoms(molecules) gives rise to an internal field at any given atom
in addition to the average macroscopic field, so that the total
fields at the atom are different from the macroscopic fields[24].In
order to achieve the negative permittivity and permeability,here the
chosen vapor with atomic concentration$N=5\times10^{16}cm^{-3}$
should be dense,so that one should consider the local field
effect,which results from the dipole-dipole interaction between
neighboring atoms.In what follows we first obtain the atomic
electric and magnetic polarizabilities, and then consider the local
field correction to the electric and magnetic susceptibilities(and
hence to the permittivity and permeability)of the coherent vapor
medium. With the formula of the atomic electric polarizations
$\gamma_{e}=2d_{43}\rho_{34}/\epsilon_{0}E_{p}$,where$E_{p}=\hbar\Omega_{p}/d_{43}$
one can arrive at
\begin{eqnarray}
\gamma_{e}=\frac{2d_{43}^2\rho_{34}}{\epsilon_{0}\hbar\Omega_{p}}\
\end{eqnarray}
In the similar fashion, by using the formulae of the atomic magnetic
polarizations $\gamma_{m}=2\mu_{0}\mu_{42}\rho_{24}/B_{p}$ [23],and
the relation of between the microscopic local electric and magnetic
fields $E_{p}/B_{p}=c$ we can obtain the explicit expression for the
atomic magnetic polarizability.Where $\mu_{0}$is the permeability of
vacuum,c is the speed of light in vacuum.Then,we have obtained the
microscopic physical quantities $\gamma_{e}$ and $\gamma_{m}$ .In
order to achieve a significant magnetic response,it should be noted
that the transition frequency between levels $|3\rangle$-$|4\rangle$
,and $|2\rangle$-$|4\rangle$ should be approximately equal to the
frequency of the probe light.Thus,the coherence $ \rho_{34}$ drives
an electric dipole,while the coherence $\rho_{24}$ drives a magnetic
dipole. However,what we are interested in is the macroscopic
physical quantities such as the electric and magnetic
susceptibilities which are the electric permittivity and magnetic
permeability.The electric and magnetic Clausius-Mossotti relations
can reveal the connection between the macroscopic and microscopic
quantities. According to the Clausius-Mossotti relation [23],one can
obtain the electric susceptibility of the atomic vapor medium
\begin{eqnarray}
\chi_{e}=N\gamma_{e}\cdot{{{{(1-\frac{N\gamma_{e}}{3})}}}}^{-1}
\end{eqnarray}
The relative electric permittivity of the atomic medium reads
$\varepsilon_{r}=1+\chi_{e}$.In the meanwhile,the magnetic
Clausius-Mossotti [24]
\begin{eqnarray}
\gamma_{m}=\frac{1}{N}(\frac{\mu_{r}-1}{\frac{2}{3}+\frac{\mu_{r}}{3}})
\end{eqnarray}
shows the connection between the macroscopic magnetic permeability
$\mu_{r}$ and the microscopic magnetic polarizations $\gamma_{m}$.It
follows that the relative magnetic permeability of the atomic vapor
medium is
\begin{eqnarray}
\mu_{r}=\frac{1+\frac{2}{3}N\gamma_{m}}{1-\frac{1}{3}N\gamma_{m}}
\end{eqnarray}
In the above,we obtained the expressions for the electric
permittivity and magnetic permeability of the four-level coherent
atomic vapor system.In the section that follows,we will demonstrate
that under the appropriate parameters condition the permittivity and
permeability of the atomic vapor system can be simultaneously
negative within the transparent window,and the absorption suppressed
deeply resulting from the EIT.

\section{Results and discussion}

In the following,the permittivity and the permeability can be
obtained by the stationary solutions to the density-matrix
equations(2)-(10).And several typical parameters should be selected
before the calculation.The density of atoms N was chosen to be
$5\times10^{16}cm^{-3}$.The parameters for the electric and magnetic
polarizabilities of atoms are chosen as: the electric transition
dipole moments $d_{43}$=2.5$\times$$10^{-29}$C$\cdot$m ,and the
magnetic dipole moments $\mu_{42}$=7.0 $\times$$10^{-23}$C$\cdot
m^{2}s^{-1}$[25], respectively.For simplicity,other parameters are
scaled by $\gamma=10^{6}s^{-1}$:
$\gamma_{1}$=$\gamma_{2}$=$\gamma_{3}$=
1$\gamma$,$\Delta_{1}$=$-1.5\gamma$, $\Delta_{2}$=1.5$\gamma$. The
Rabi frequency of the probe field is $\Omega_{p}=0.01\gamma$. The
strong coupling optical field drivers the atomic system with Rabi
frequencies  $|\Omega_{1,2}|$=$2.5\gamma$,and its phase
$\Phi$=$-\frac{3}{4}\pi$.

\begin{figure}[h!]
  \centering
  \includegraphics[width=3.5in]{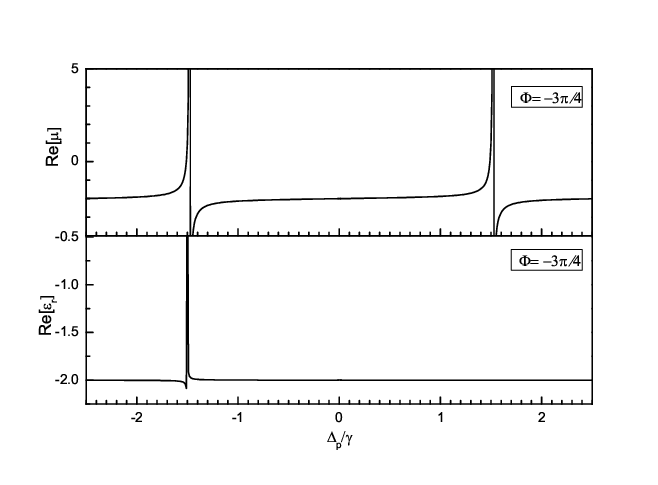}
    \includegraphics[width=3.5in]{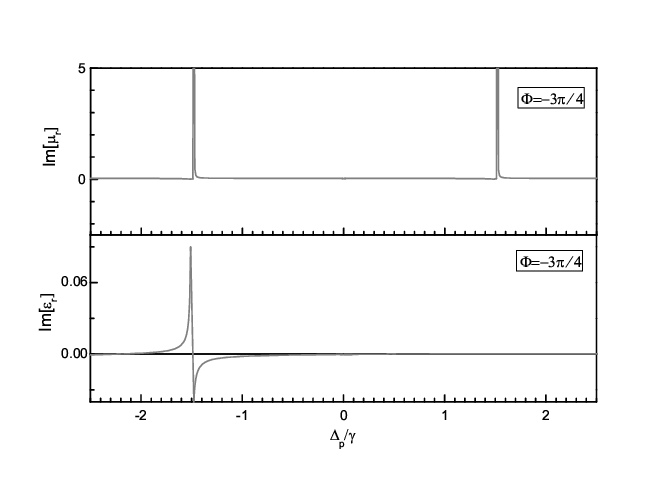}
  \hspace{0in}%
  \caption{The real and imaginary parts of the
permeability $\mu_{r}$ and permittivity $\epsilon_{r}$ as a function
of the rescaled  probe field detuning $\frac{\Delta_p}{\gamma}$,and
the other parameters given in the text.}
\hspace{0in}%
\end{figure}\label{Fig.2}

To analyze whether an atomic system has left-handedness or not,one
should consider the plus or minus of the the relative electric
permittivity $\epsilon_{r}$ and magnetic permeability $\mu_{r}$.In
Fig.2 we plot the calculated electric permittivity $\epsilon_{r}$
and magnetic permeability $\mu_{r}$ as a function of the rescaled
probe field detuning.In order to compare the simultaneous negative
value of $\epsilon_{r}$ and $\mu_{r}$ ,the real and imaginary parts
of the relative electric permittivity $\epsilon_{r}$ and magnetic
permeability $\mu_{r}$ are plotted together,respectively.From the
real parts images,it is observed that the real part of relative
electric permittivity is more convenient to be negative,and the real
part magnetic permeability $\mu_{r}$ shows negative value in the
frequency bands[$-2.5\gamma$,$-1.452\gamma$],[$-1.501\gamma$,
$1.495\gamma$] and [$1.525\gamma$,$2.5\gamma$]. So the
simultaneously negative values for them are consistent with the real
part magnetic permeability $\mu_{r}$.The atomic system has
left-handedness in frequency bands with the simultaneously negative
values. Now we turn our attention to their imaginary parts in
Fig.2.The property of their imaginary parts are be of particular
interest and importance to us.And we pay much attention to the
frequency bands with the simultaneously negative values for their
real parts.And both their imaginary parts show clearly transparent
propagation in the same frequency bands.This is a very significant
result for us. As is well known, the photon absorption of atom can
be greatly suppressed via electromagnetically induced transparency
(EIT)[22].In the transparent area,we may predict low absorption
obtained by the EIT phenomenon in the left-handed atomic system.

\begin{figure}[h!]
  \centering
  \includegraphics[width=3.5in]{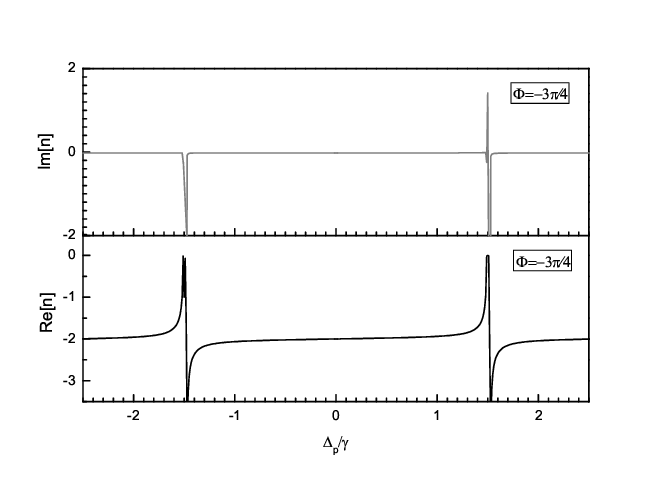}
  \hspace{0in}%
  \caption{The refractive index as a function of the
rescaled  probe field detuning $\frac{\Delta_p}{\gamma}$,and the
other parameters given in the text.}
\hspace{0in}%
\end{figure}\label{Fig.3}

The refractive index according to definition of the left-handed
material ($n(\omega)$=
$-\sqrt{\epsilon_{r}(\omega)\cdot\mu_{r}(\omega)}$) [1]is plotted as
a function of the rescaled probe field detuning
$\frac{\Delta_p}{\gamma}$ in Fig.3.As shown in Fig.3, we notice that
the atomic system gets the value approximating to -2 as its
refractive index in frequency bands for $\epsilon_{r}$ and $\mu_{r}$
having simultaneous negative values.In the same frequency bands,its
imaginary part depicts the stirring property.With observation of the
profile,we note that a broad transparent area in the curve.It
demonstrates that the atomic vapor system displays negative
refraction with very low absorption in the region.Now we provide a
qualitative explanation for the above numerical results.It can be
seen from Fig.2 that both \emph{Im}[$\epsilon_{r}$] and
\emph{Im}[$\mu_{r}$]show clearly transparent propagation in the same
frequency bands.During the calculation of
$n(\omega)$=$-\sqrt{\epsilon_{r}(\omega)\cdot\mu_{r}(\omega)}$,the
result of the imaginary part of refractive index has to be
transparent.Thus, this demonstrates the low absorption resulting
from the EIT as shown in Fig.2 .

\begin{figure}[h!]
  \centering
  \includegraphics[width=3.5in]{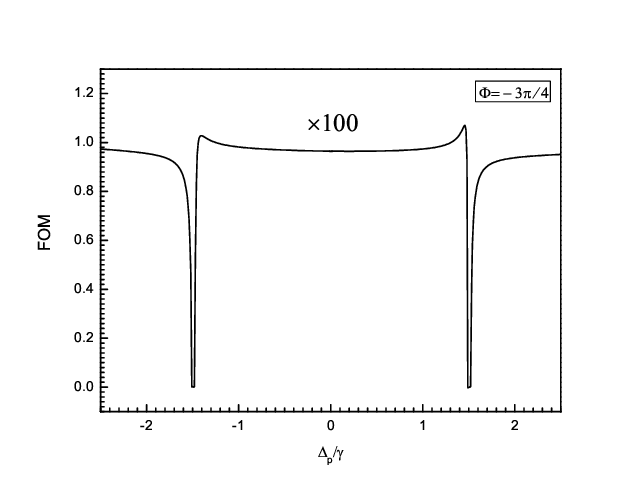}
  \hspace{0in}%
  \caption{Figure of merit (FOM) as a function of the rescaled  probe
field detuning $\frac{\Delta_p}{\gamma}$,and the other parameters
given in the text.}
\hspace{0in}%
\end{figure}\label{Fig.4}

Since the performance of the left-handed materials is typically
characterized by the figure of merit.We plot the figure of merit
(FOM)in Fig.4.In the frequency region of simultaneous negative for
$\epsilon_{r}$ and $\mu_{r}$,the negative refractive index appears
as shown in Fig.3, and the FOM is much larger than unity. It means
that there is almost no absorption in this area.By comparison with
various artificial structures in the realization of negative
refraction, the present scheme is featured by the absorption
depression to a very low level.According to the results shown in
Fig.2,one may have one reasons for the absorption depression.
Because of the EIT effect, there is an electric non-absorption (
$\epsilon_{r}$=$\emph{Re}$[$\epsilon_{r}$]+$\textbf{\emph{i}}$$\cdot$$\emph{Im}$[$\epsilon_{r}$],$\emph{Im}$[$\epsilon_{r}$]$\approx$0)
in the frequency region of simultaneous negative for $\epsilon_{r}$
and $\mu_{r}$.At the same time, the magnetic response of the medium
is($\mu_{r}$=$\emph{Re}$[$\mu_{r}$]+$\textbf{\emph{i}}$$\cdot$$\emph{Im}$[$\mu_{r}$],$\emph{Im}$[$\mu_{r}$]$\approx$0)in
the same frequency region. Thus,the imaginary part of
$n$=$-\sqrt{\epsilon_{r}\cdot\mu_{r}}$=$-\sqrt{(\emph{Re}[\epsilon_{r}]+i\cdot\emph{Im}[\epsilon_{r}])\cdot(\emph{Re}[\mu_{r}]+i\cdot\emph{Im}[\mu_{r}])}$
is is very small,and the ratio of its real and imaginary parts is
over 100 in the same frequency regions, which means the low
absorption in the frequency regions.

\section{Conclusion}

In summary, we have demonstrated a scheme for realizing negative
refractive index with low absorption in a left-handed atomic
system.The simultaneously negative permittivity and permeability are
obtained,and the transparent propagation shows the EIT effect in the
four-level atomic system. Due to the EIT effect can inhibit the
absorption, naturally, the atomic system obtains low absorption of
negative refraction in our conclusion.Therefore,our aim for
searching the low-loss negative refraction is achieved under the
choosing appropriate parameters in resonant atomic system,in respect
that the main applied limitation of the negative refractive
materials is the large amount of dissipation and absorption.
\section*{Acknowledgments}
The work is supported by the National Natural Science Foundation of
China ( Grant No.60768001 and No.10464002 ).

\end{document}